\documentclass[twocolumn]{aastex701}
\usepackage[table]{xcolor}
\usepackage{hyperref}
\usepackage{amsmath}
\usepackage{amssymb}
\usepackage{float}
\usepackage{subcaption}
\usepackage{graphicx}
\usepackage{placeins}
\usepackage{multirow}
\usepackage{colortbl}
\usepackage{hhline}
\usepackage{csquotes}
\usepackage{todonotes}

\newcommand{\vect}[1]{\mathbf{\boldsymbol{#1}}}

\newcommand{\chieff}{\chi_\text{eff}}

\newcommand{\rateunit}{\text{Gpc}^{-3}\text{yr}^{-1}}

\begin{document}

\title{The First Detection of Sub-Populations in the Delay-Time Distribution of Binary Black Holes in GWTC-4 of LIGO-Virgo-KAGRA}

\author[orcid=0009-0006-3960-9405,gname=Shaunak, sname='Padhyegurjar']{Shaunak Padhyegurjar}
\affiliation{Department of Astronomy and Astrophysics, Tata Institute of Fundamental Research, Homi Bhabha Road, Navy Nagar, Colaba, Mumbai 400005, India}
\email[show]{shaunak.padhyegurjar@tifr.res.in}  

\author[orcid=0000-0002-3373-5236,gname=Suvodip, sname='Mukherjee']{Suvodip Mukherjee} 
\affiliation{Department of Astronomy and Astrophysics, Tata Institute of Fundamental Research, Homi Bhabha Road, Navy Nagar, Colaba, Mumbai 400005, India}
\email[show]{suvodip@tifr.res.in}

\begin{abstract}
The imprint of different formation channels of binary black holes (BBHs) is encoded in the distribution of time delays between BBH mergers and the formation of their progenitor stars, along with their source properties such as component mass, mass-ratio, spin, and more. This makes it possible for the presence of a potential correlation between the delay-time distribution and compact-object source properties. We report the first measurement of this inevitable signature from the fourth gravitational wave (GW) catalog (GWTC-4) of LIGO-Virgo-KAGRA and identified three sub-populations that show distinct merger rate behavior as a consequence of this. We find that the delay-time distribution of the sources above a mass of $45$ M$_\odot$ is significantly different from the ones below and exhibits strong dependence on the mass-ratio and spin, indicating that GW sources close to equal masses and close to zero effective spin are more delayed in comparison to the values otherwise. Our analysis identifies the presence of at least three source property dependent sub-population of merger rates with the merger rate at redshift $z=0$ varying from $\sim 0.6- 12$ Gpc$^{-3}$ yr$^{-1}$ for the three different sub-populations and hence rule out a Universal merger rate for all the BBHs detected using GW.

\end{abstract}

\section{Introduction}

The redshift distribution of binary black holes (BBHs) in the Universe and its evolution with cosmic time remains an intriguing scientific question for decades, which is now beginning to be answered with the help of gravitational wave (GW) observations made by the LIGO-Virgo-KAGRA (LVK) collaboration with the advanced LIGO, Virgo, and KAGRA detectors \citep{LIGOScientific:2014pky, VIRGO:2014yos, Virgo:2019juy, Virgo:2022ysc, KAGRA:2013rdx, Aso:2013eba, KAGRA:2020tym}. With the discovery of nearly 150 BBH events until the fourth Gravitational Wave Transient Catalog (GWTC-4) \citep{LIGOScientific:2025slb}, GW observations have begun to reveal the statistical properties of the BBH population, including their mass spectrum, spin distribution, and merger rate evolution. This progress is yet to reveal the complete understanding of the BBH population. Most notably, the formation channels responsible for producing the observed BBH population remain an open question.  One of the smoking gun signatures of different formation channels is the delay-time distribution (DTD), which characterizes the time delay between the formation of the progenitor system and the BBH merger. Since different astrophysical environments and evolution pathways are expected to produce different DTDs, measurements thereof can provide important clues about the origin of merging BBHs.

\begin{figure*}
    \centering
    \includegraphics[scale=0.7]{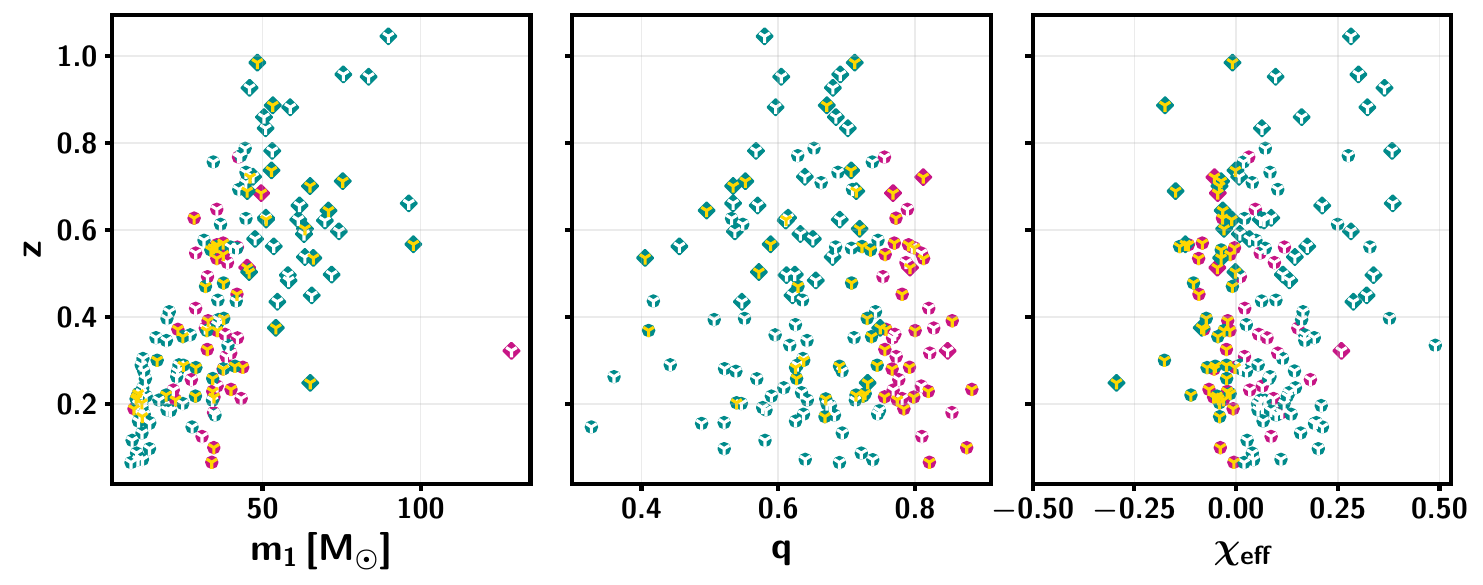}
    \caption{Distribution of the observed properties of GWTC-4 BBH events in primary mass ($m_1$), mass-ratio ($q$), effective spin ($\chieff$), and redshift ($z$). Sources with $m_1>\text{M}_\text{PISN}$ are marked by \emph{diamonds}, while \emph{circles} represent $m_1<\text{M}_\text{PISN}$. Sources with $q>0.75$ are shown in \emph{violet-red}, whereas sources with $q<0.75$ are shown in \emph{cyan}. In addition, \emph{white triads} indicate sources with $\chieff>0$, while \emph{gold triads} correspond to sources with $\chieff<0$.}
    \label{fig:data}
\end{figure*}

The distribution of delay time of the binaries depends on the channels through which they form, and also the initial distribution of the separation of the parent stars. The vanilla model usually considered is a power-law form of the delay-time distribution (DTD) $p(t) \propto t^{-d} \mathrm{\,for\,} t\geq t_{\rm min}$, with $d$ denoting the power-law index and $t_{\rm min}$ denoting the minimum value of the DTD. Previous analyses from GW data have constrained these quantities with a consensus that large values of delay-time are not possible \citep{Mukherjee:2021ags, Fishbach:2021mhp, Karathanasis:2022rtr,Vijaykumar:2023bgs, Turbang:2023tjk}. There are also previous analyses showing the requirements of going beyond the power-law form of the DTD \citep{Afroz:2025typ}. However, the data does not seem to provide any strong evidence towards the requirement for going beyond a power-law form to explain the GWTC-4 BBH sources \citep{Fishbach:2018edt, Safarzadeh:2020qru, Fishbach:2021mhp, Mukherjee:2021rtw, Karathanasis:2022rtr, Smith:2024awx}. But there remains an interesting unanswered question, \textbf{Are there multiple populations of  BBHs present in the GW catalog which are originating from different DTDs depending on their properties?}

In this work, we address this question for the first time by carrying out a hierarchical Bayesian analysis using the code \texttt{BBH-Genesis} \citep{bbhgenesis} on GWTC-4 for BBHs depending on their source properties, namely primary masses and effective spin parameters, which are relatively well measured from individual events. We show the properties of the observed GW sources in the GWTC-4 catalog in Fig. \ref{fig:data}. Our analysis discovers the existence of three sub-populations of the DTD resulting in three distinct local merger rate and redshift evolution of the BBHs. We find that the DTD and local merger-rate of BBHs with masses below the lower edge of pair instability of supernovae (PISN) M$_{\rm PISN} \sim 45$ M$_\odot$ are significantly different from the DTD and local merger-rate of sources above M$_{\rm PISN}$. Moreover, sources with primary mass above M$_{\rm PISN}$ exhibit two different sub-populations of delay-time and merger rate for mass ratio $q \equiv m_2/m_1$ greater than or less than 0.75. This finding stands even for different models of metallicity-dependent star formation rate (SFR). As a result, this finding discovers for the first time the existence of three different sub-populations of DTD in the data in a data-driven way, without assuming any specific formation channels. This work provides the observational evidence towards the previous theoretical predictions of the dependence of DTD on GW sources \citep{Mandel:2015qlu,Boesky:2024wks,2025A&A...698A.144S}. In the future, bridging these findings with a specific formation channel model can shed light on the formation channels of BBHs in the Universe and the relative dominance of one over the other. 

This paper is organized as follows: we explain the method used in the analysis in Sec. \ref{sec:methods}. Followed by that, we explain the results and discussion in Sec. \ref{sec:results} and \ref{sec:discussion} respectively. Finally, in Sec. \ref{sec:conclusion} we discuss the summary of the findings and future prospects.

\section{Methods}\label{sec:methods}
\subsection{Astrophysical modeling of DTD}
The DTD of BBHs remains vastly unknown from the observational side. There are multiple theoretical studies based on stellar population analysis models that predict a power-law form of the DTD, as well as possible variations from them due to different formation channels \citep{2010ApJ...716..615O, 2016A&A...588A..50M,2017MNRAS.472.2422M, 2017ApJ...841...77A,Rodriguez:2018jqu,2018PhRvL.120o1101R, DiCarlo:2020lfa, Fragione:2021hhl}. Theoretical studies have also indicated that there can be an intrinsic dependence of DTD on the metallicity-dependent SFR in the Universe \citep{Mandel:2015qlu,Boesky:2024wks}. 

On the other hand, the source properties of BBHs, such as component masses, mass ratios, and spin parameters, also carry essential imprints of their formation channels. Sources above PISN mass scale, non-aligned spin distribution, unequal mass binaries are some of the smoking gun features which can be useful in disentangling different formation channels \citep{Mandel:2018hfr,2020ApJ...894..133A, Mapelli:2021taw}. As a result, it is likely to be a natural consequence that the imprint of different formation channels of BBHs will lead to different DTDs, which will also depend on the BBH source properties, such as its component masses, mass ratio, spin, etc. Though such predictions from the theoretical side remain susceptible to modeling uncertainties and simplified assumptions, current data can give a hint of its existence for searches that are not specific to any formation channel (and hence model-independent).

\begin{figure}
    \centering
    \includegraphics[scale=0.55]{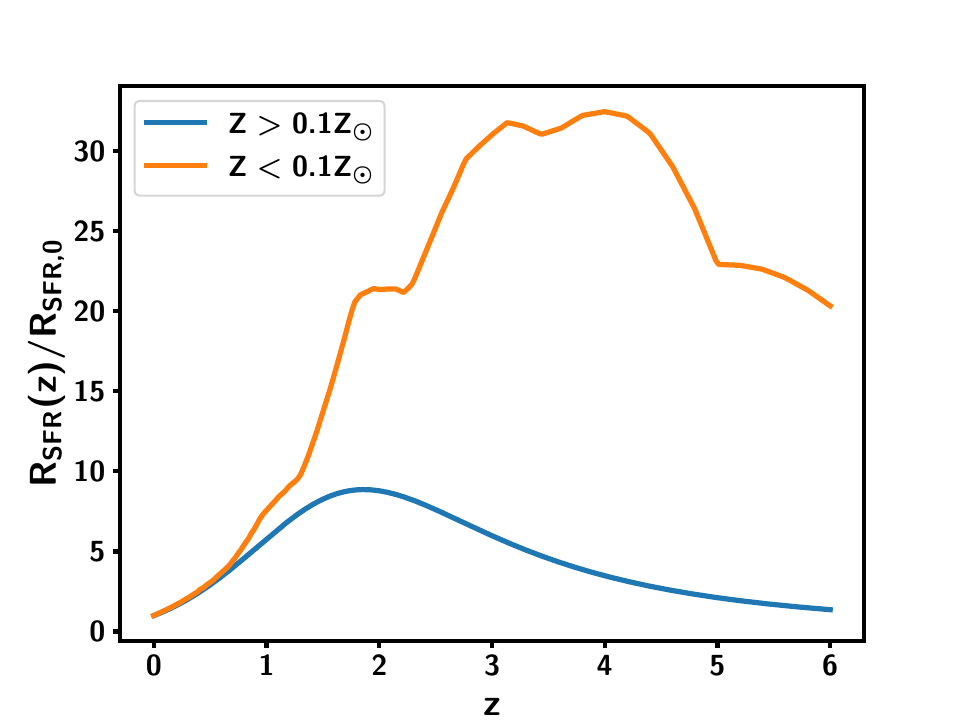}
    \caption{Comparison of the evolution of the normalized SFR ($R_{\rm SFR}(z)/R_{\rm SFR}(z=0)$) as a function of redshift $z$ for two different cases of metallicity -- high-$Z$ ($Z>0.1Z_\odot$, shown in \emph{blue}) and low-$Z$ ($Z<0.1Z_\odot$, shown in \emph{orange}).}
    \label{fig:sfrcomparison}
\end{figure}

Based on the philosophy mentioned above, we consider a model-independent data-driven approach to find out whether there is a classification of the DTD of the BBH merger rate depending on the source properties. In order to do this, we write down a merger rate model of the BBHs as a function of source properties and metallicity as
\begin{widetext}
\begin{align*}
    \psi(Z,z,\{\Theta^i_{\rm GW}\}) = \frac{\int_z^\infty p_t(t_d|d (\{\Theta^i_{\rm GW}\}), t_d^\text{min} (\{\Theta^i_{\rm GW}\}), t_d^\text{max}(\{\Theta^i_{\rm GW}\})) R_\text{SFR}(Z,z_f) \frac{dt}{dz_f}dz_f}{\int_0^\infty p_t(t_d|d (\{\Theta^i_{\rm GW}\}), t_d^\text{min} (\{\Theta^i_{\rm GW}\}), t_d^\text{max} (\{\Theta^i_{\rm GW}\})) R_\text{SFR}(Z,z_f) \frac{dt}{dz_f}dz_f}.
\end{align*}
\end{widetext}
where $\{\Theta^i_{\rm GW}\}$ denotes the GW source properties such as component masses, mass-ratio, and spin, and the time delay distribution model can depend on these parameters for a parametric form given by \citep{Dominik:2014yma, Mandel:2015qlu, Vitale:2018yhm, Fishbach:2021mhp, Mukherjee:2021rtw, Karathanasis:2022hrb, Karathanasis:2022rtr}
\begin{widetext}
\begin{align}
    p_t(t_d|d (\{\Theta_{\rm GW}\}), t_d^{\text{min}} (\{\Theta_{\rm GW}\}), t_d^{\text{max}} (\{\Theta_{\rm GW}\})) \propto \begin{cases}
        (t_d)^{-d(\{\Theta_{\rm GW}\})} & \quad ,\,t_d^{\text{min}}(\{\Theta_{\rm GW}\}) < t_d < t_d^\text{max} ((\{\Theta_{\rm GW}\})\\
        0 & \quad ,\,\text{otherwise}
    \end{cases},
\end{align}
\end{widetext}
where time delay defined as $t_d=t(z_m)-t(z_f)$, with $t(z)$ being the age of the universe at redshift $z$, and $z_f$ and $z_m$ being the redshifts of formation and merger respectively.  The metallicity-dependent SFR remains a challenging quantity to infer from astrophysical observations. In this analysis, we have used the metallicity-dependent SFR from \citep{2021MNRAS.508.4994C}. At the higher metallicity values, it matches well with the Madau-Dickinson SFR model $R_\text{SFR}(Z>0.1Z_\odot, z)$  \citep{Madau:2014bja}, denoted by $Z>\,0.1Z_\odot$ shown in Fig. \ref{fig:sfrcomparison}\footnote{$Z_\odot$ denotes the solar metallicity.}. 

For the low-metallicity values $Z<0.1Z_\odot$, the metallicity distribution differs from the Madau-Dickinson SFR. It primarily peaks at a higher redshift than redshift $z=2$. This is expected as the high-redshift Universe is primarily driven by the low-metallicity stars. We suggest that the readers read the papers from \citet{Chruslinska:2018hrb, Chruslinska:2022ovf}  for more details behind the metallicity-dependent SFR model used in this analysis.

To understand the dependence of the DTD on the BBH source properties $\{\Theta_{\rm GW}\}$, we consider a broad classification, namely:
\begin{enumerate}
    \item Sources with primary mass above M$_{\rm PISN}$ ($m_1>\rm{M}_{\rm PISN}$).
    \begin{itemize}
        \item $\chieff> 0$ or $\chieff<0$ (denoted by $\{\Theta^1_{\rm GW}\}$).
        \item $q>\bar q$ or $q<\bar q$ (denoted by $\{\Theta^2_{\rm GW}\}$).
    \end{itemize}
     \item Sources with primary mass below M$_{\rm PISN}$ ($m_1<\rm{M}_{\rm PISN}$).
     \begin{itemize}
        \item $\chieff> 0$ or $\chieff<0$ (denoted by $\{\Theta^3_{\rm GW}\}$).
        \item $q>\bar q$ or $q<\bar q$ (denoted by $\{\Theta^4_{\rm GW}\}$).
    \end{itemize}
\end{enumerate}
For this analysis, we have chosen $\bar q=0.75$. This choice is to primarily demarcate between sources arising from formation channels that support equal-mass binaries, such as the isolated formation channel, and other channels. Similarly, for $\chieff$, this choice is made to make a broad classification in this analysis. Results for a few different classifications are shown in Appendix \ref{app:chi}. In principle, one can look for possible parametric dependence of the time-delay distribution on the GW source parameters. We defer this for future work. It is important to note here that the dependence of M$_{\rm PISN}$ on parent star metallicity is a crucial physical effect, which can produce black holes of masses above M$_{\rm PISN}= 45$ M$_\odot$. As a result, for the low-metallicity SFR ($Z < 0.1\, Z_\odot$), we perform the analysis only for sources with primary masses above 45  M$_\odot$.

\subsection{Hierarchical Bayesian Analysis}
We infer the properties of the observed population of BBHs using hierarchical Bayesian inference \citep{Loredo:2004nn, ThraneTalbot2019, Vitale:2020aaz} to estimate the hyperparameter posterior $p(\vect{\Lambda}|\vect{\mathcal{D}})$ using data of $N_\text{det}$ observed GW events $\vect{\mathcal{D}}=\{\vect{d}_1,\ldots,\vect{d}_{N_\text{det}}\}$ by using the likelihood of the form \citep{LIGOScientific:2025pvj}

\begin{align}
    \mathcal{L}(\vect{\mathcal{D}}|\vect{\Lambda}) \propto \prod_{i=1}^{N_\text{det}}\frac{\int d\vect{\theta}\mathcal{L}(\vect{d}_i|\vect{\theta})\pi(\vect{\theta}|\vect{\Lambda})}{\xi(\vect{\Lambda})},\label{eqn:poplike}
\end{align}
where $\mathcal{L}(\vect{d}_i|\vect{\theta})$ are the likelihoods of individual events and $\xi(\vect{\Lambda})$ is the selection function given by
\begin{align}
    \xi(\vect{\Lambda}) = \int_{\rho(\vect{d})>\rho_\text{th}}d\vect{d}d\vect{\theta}\mathcal{L}(\vect{d}|\vect{\theta})\pi(\vect{\theta}|\vect{\Lambda}).\label{eqn:selfn}
\end{align}

\begingroup
\begin{table*}
    \centering
    \setlength{\tabcolsep}{3pt}
    \begin{tabular}{cccccccccccccccccccc}
        \hline\hline
        $\alpha_1$ & $\alpha_2$ & $m_\text{break}$ & $\mu_1$ & $\sigma_1$ & $\mu_2$ & $\sigma_2$ & $m_\text{low,1}$ & $\delta_{m,1}$ & $m_\text{max}$ & $\lambda_0$ & $\lambda_1$ & $\beta$ & $m_\text{low,2}$ & $\delta_{m,2}$ & $\mu_\chi$ & $\sigma_\chi$ & $\zeta$ & $\mu_t$ & $\sigma_t$ \\
        \hline
        1.57 & 4.61 & 36.02 & 9.76 & 0.64 & 32.55 & 4.28 & 4.94 & 3.73 & 300 & 0.36 & 0.59 & 1.57 & 3.54 & 4.59 & 0.08 & 0.32 & 0.65 & 0.48 & 1.23 \\
        \hline
    \end{tabular}
    \caption{Fixed values of the population parameters such as primary mass, mass ratio, and spin distribution hyperparameters.}
    \label{tab:bestfit_mass_spin}
\end{table*}
\endgroup

The selection function corrects for the Malmquist bias that arises from the selective efficiency of GW detectors to observe events that pass the detection threshold $\rho_\text{th}$ in the detection statistic $\rho(\vect{d})$ (SNR). Based on the \texttt{BBH-Genesis} \citep{bbhgenesis} code developed, we factorize the distribution as 
\begin{widetext}
    \begin{align}
        \pi(m_1, q, a_1, a_2, \cos\theta_1, \cos\theta_2, z|\vect{\Lambda}) = \pi(m_1|\vect{\Lambda})\pi(q|m_1, \vect{\Lambda}) \pi(a_1, a_2, \cos\theta_1, \cos\theta_2| \vect{\Lambda})\pi(z|\vect{\Lambda}). \label{eqn:chmodel}
    \end{align}
\end{widetext}
The redshift distribution for each case is
\begin{align}
    \pi(z|\vect{\Lambda}) = \left(\frac{dV_c}{dz}\frac{1}{1+z} \right) R_0(\{\Theta^i_{\rm GW}\})\psi(Z,z,\{\Theta^i_{\rm GW}\}|\vect{\Lambda}),\label{eqn:chredshift}
\end{align}
where $R_0(\{\Theta^i_{\rm GW}\})$ is the local merger rate (redshift $z=0$) and $\psi(Z,z=0,\{\Theta^i_{\rm GW}\}|\vect{\Lambda}) = 1$ for different cases considered in the analysis with $i \in \{1, 2, 3,4\}$. 

We use the default models used in \citep{LIGOScientific:2025pvj} for the primary mass, mass ratio, and spin distributions, with \textsc{Broken Power Law + Two Peaks} model for $p(m_1)$, a power law model for $p(q)$, and \textsc{Gaussian Component Spins} model for the spin distribution. Since our analysis focuses only on the redshift distribution for different cases, we do not simultaneously vary the mass and spin hyperparameters in each run. Instead, we fix these distributions using the best-fit values of hyperparameters obtained from a joint inference of the primary mass, mass ratio, spin distributions, and the time delay redshift model for the high-$Z$ SFR for the whole sample of events considered in this analysis. The best fit values of hyperparameters in primary mass, mass ratio, and spin distributions are given in Table \ref{tab:bestfit_mass_spin}. Note that for the cases involving cuts on $\chieff$ values of events, the corresponding values of spin components enter the above spin distribution. The DTD inferred jointly with the entire population parameters for different population models is also shown in a companion paper \citep{bbhgenesis}. All results are quoted at 90\% credible intervals. 

\section{Results}\label{sec:results}

\begin{figure}
    \centering
    \begin{minipage}{0.49\textwidth}
        \centering
        \includegraphics[scale=0.475]{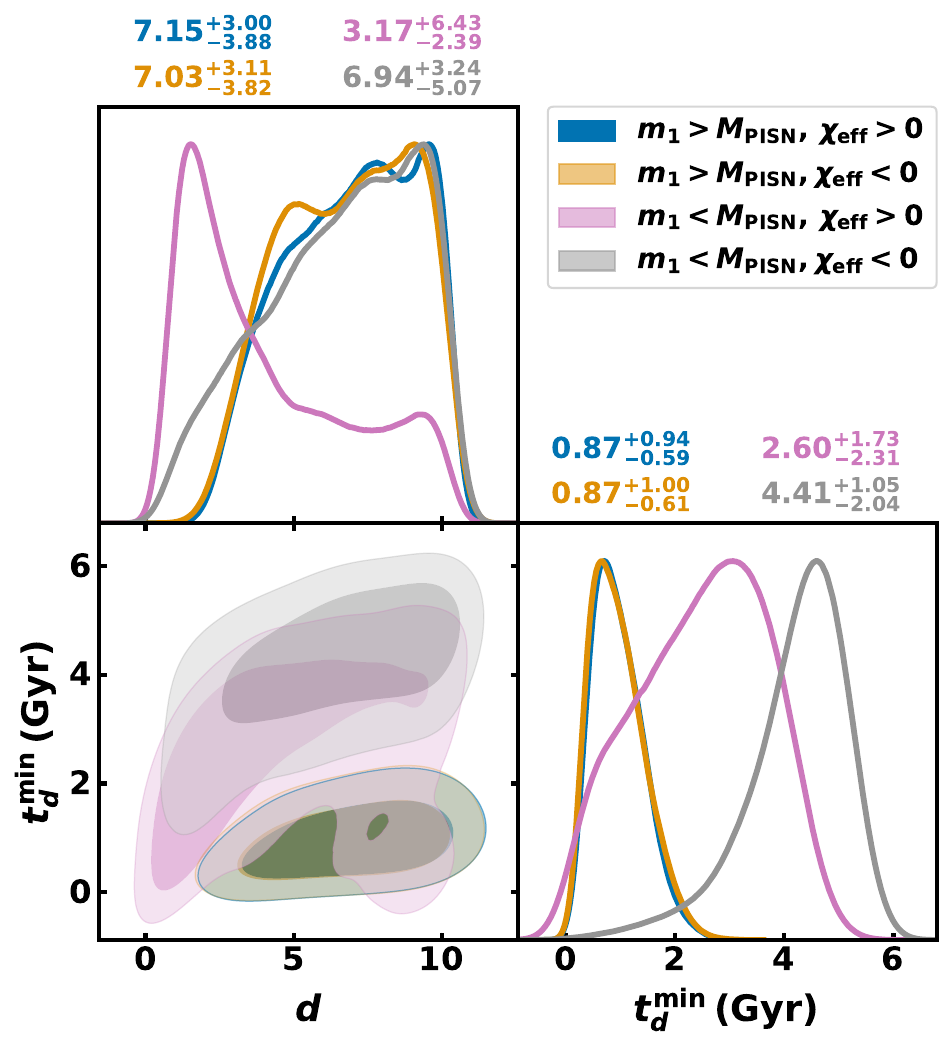}
        \end{minipage}
        \begin{minipage}{0.49\textwidth}
        \centering
        \includegraphics[scale=0.475]{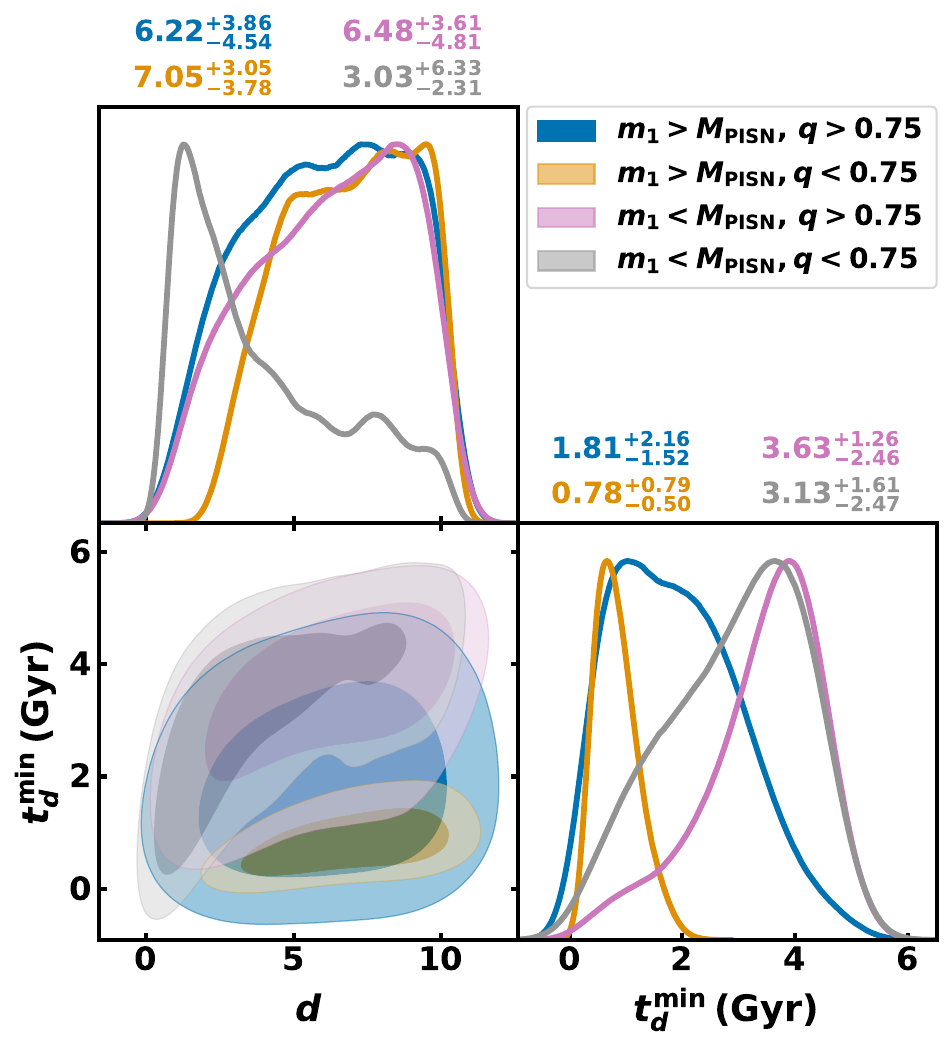}
        \end{minipage}
    \caption{Inferred values of $t_d^\text{min}$ and $d$ for high-$Z$ SFR for the cases: (\emph{Top}) $\chieff>0$ and $\chieff<0$ for $m_1>\text{M}_\text{PISN}$ ($\{\Theta^1_{\text{GW}}\}$) and the same for $m_1<\text{M}_\text{PISN}$ ($\{\Theta^3_{\text{GW}}\}$). (\emph{Bottom}): $q>0.75$ and $\chieff<0$ for $m_1>\text{M}_{\text PISN}$ ($\{\Theta^2_{\text{GW}}\}$) and the same for ($m_1<\text{M}_{\text PISN}$) ($\{\Theta^4_{\text{GW}}\}$).}\label{fig:corner-1}
\end{figure}

\begin{figure*}
    \centering
    \begin{minipage}{0.49\textwidth}
        \centering
        \includegraphics[scale=0.55]{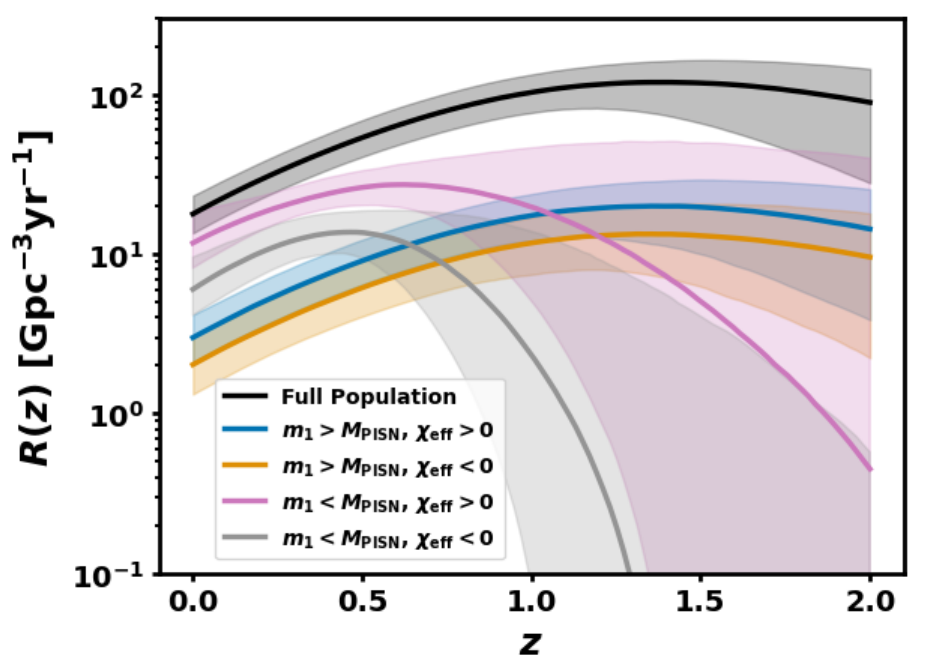}
    \end{minipage}
    \begin{minipage}{0.49\textwidth}
        \centering
        \includegraphics[scale=0.55]{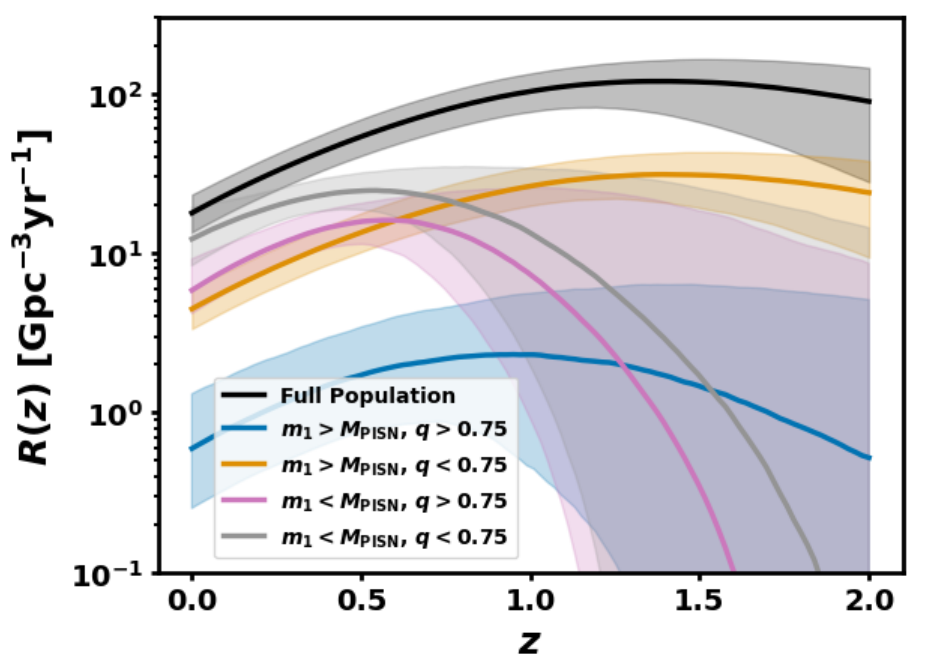}
    \end{minipage}
    \caption{Redshift distributions inferred with high-$Z$ SFR for the cases: (\emph{Left}) $\chieff>0$ and $\chieff<0$ for $m_1>\text{M}_\text{PISN}$ ($\{\Theta^1_{\text{GW}}\}$) and the same for $m_1<\text{M}_\text{PISN}$ ($\{\Theta^3_{\text{GW}}\}$). (\emph{Right}): $q>0.75$ and $\chieff<0$ for $m_1>\text{M}_{\text PISN}$ ($\{\Theta^2_{\text{GW}}\}$) and the same for ($m_1<\text{M}_{\text PISN}$) ($\{\Theta^4_{\text{GW}}\}$). Also shown in black the inferred distribution for the entire population.}\label{fig:chieff_highZ}
\end{figure*}

We perform the analysis on GWTC-4 data publicly released by the LVK Collaboration \citep{gwtc4v2, gwtc3v2, gwtc2v2} for 153 BBH events with FAR $\leq1\,\text{yr}^{-1}$. We use the \textsc{NrSur7dq4} posterior samples for events in GWTC-4 when available\footnote{If event posterior samples are not available for the waveform model \textsc{NrSur7dq4}, then we use \textsc{Mixed} posterior samples for our analysis.} to minimize any bias in hierarchical population inference resulting from waveform systematics \citep{Das:2026glo}.  We correct for selection effects using the suite of injections publicly released alongside GWTC-4 \citep{Essick:2025zed, sensInj}. We also impose a maximum threshold in likelihood uncertainty estimate $\sigma^2_{\ln\mathcal{L}}<1$ to manage uncertainty arising from using a finite number of samples in the Monte Carlo estimation of integrals in the likelihood (see Eq. \eqref{eqn:poplike}).

{
\setlength{\tabcolsep}{3pt}
\begin{table*}
\centering
\renewcommand{\arraystretch}{1.15}

\begin{tabular}{|c|c|c|c|c|c|c|c|c|c|c|}
\hline
\multirow{3}{*}{\textbf{Case}}
& \multirow{3}{*}{\textbf{Cut}}
& \multirow{3}{*}{\shortstack[c]{\textbf{No. of}\\ \textbf{Events}}}
& \multicolumn{8}{c|}{\textbf{Estimates (90\% CI)}} \\
\hhline{|~|~|~|-|-|-|-|-|-|-|-|}
&&
& \multicolumn{2}{c|}{$R_0$ $[\rateunit]$}
& \multicolumn{2}{c|}{$t_d^\text{min}$ [Gyr]}
& \multicolumn{2}{c|}{$d$} & \multicolumn{2}{c|}{$N_\text{est}$} \\
\hhline{|~|~|~|-|-|-|-|-|-|-|-|}
&&
& High-$Z$ & Low-$Z$
& High-$Z$ & Low-$Z$
& High-$Z$ & Low-$Z$
& High-$Z$ & Low-$Z$ \\
\hline
\multirow{4}{*}{$m_1>M_\text{PISN}$}
& $\chieff>0$ & 25 & $3.00^{+1.16}_{-0.90}$ & $8.66^{+3.40}_{-2.72}$ & $0.87^{+0.94}_{-0.59}$ & $2.02^{+1.07}_{-1.25}$ & $7.15^{+3.00}_{-3.88}$ & $7.98^{+2.24}_{-3.81}$ & \cellcolor{gray!25}{$24.51^{+8.85}_{-7.07}$} & \cellcolor{gray!25}{$24.67^{+9.22}_{-7.51}$} \\
\hhline{|~|-|-|-|-|-|-|-|-|-|-|}
& $\chieff<0$ & 17 & $2.03^{+0.98}_{-0.71}$ & $5.90^{+2.94}_{-2.08}$ & $0.87^{+1.00}_{-0.61}$ & $1.89^{+1.12}_{-1.28}$ & $7.03^{+3.11}_{-3.82}$ & $7.65^{+2.54}_{-3.67}$ & \cellcolor{gray!25}{$16.58^{+7.53}_{-5.57}$} & \cellcolor{gray!25}{$16.63^{+7.60}_{-5.84}$} \\
\hhline{|~|-|-|-|-|-|-|-|-|-|-|}
& $q>0.75$ & 5 & $0.60^{+0.71}_{-0.35}$ & $1.65^{+1.75}_{-0.96}$ & $1.81^{+2.16}_{-1.52}$ & $2.70^{+2.31}_{-2.33}$ & $6.22^{+3.86}_{-4.54}$ & $6.73^{+3.39}_{-4.60}$ & \cellcolor{gray!25}{$4.58^{+4.37}_{-2.55}$} & \cellcolor{gray!25}{$4.61^{+4.38}_{-2.65}$} \\
\hhline{|~|-|-|-|-|-|-|-|-|-|-|}
& $q<0.75$ & 37 & $4.47^{+1.50}_{-1.12}$ & $12.74^{+4.07}_{-3.34}$ & $0.78^{+0.79}_{-0.50}$ & $2.08^{+0.76}_{-0.99}$ & $7.05^{+3.05}_{-3.78}$ & $8.27^{+1.99}_{-3.30}$ & \cellcolor{gray!25}{$36.50^{+11.53}_{-8.83}$} & \cellcolor{gray!25}{$36.51^{+10.93}_{-8.92}$} \\

\hline

\multirow{4}{*}{$m_1<M_\text{PISN}$}
& $\chieff>0$ & 73 & $11.74^{+6.75}_{-3.58}$ & - & $2.60^{+1.73}_{-2.31}$ & - & $3.17^{+6.43}_{-2.39}$ & - & \cellcolor{gray!25}{$72.69^{+14.35}_{-12.85}$} & - \\
\hhline{|~|-|-|-|-|-|-|-|-|-|-|}
& $\chieff<0$ & 38 & $6.02^{+3.53}_{-1.89}$ & - & $4.41^{+1.05}_{-2.04}$ & - & $6.94^{+3.24}_{-5.07}$ & - & \cellcolor{gray!25}{$37.56^{+11.51}_{-8.94}$} & - \\
\hhline{|~|-|-|-|-|-|-|-|-|-|-|}
& $q>0.75$ & 41 & $5.85^{+3.33}_{-1.69}$ & - & $3.63^{+1.26}_{-2.46}$ & - & $6.48^{+3.61}_{-4.81}$ & - & \cellcolor{gray!25}{$41.01^{+10.95}_{-9.72}$} & - \\
\hhline{|~|-|-|-|-|-|-|-|-|-|-|}
& $q<0.75$ & 70 & $12.27^{+7.02}_{-3.94}$ & - & $3.13^{+1.61}_{-2.47}$ & - & $3.03^{+6.33}_{-2.31}$ & - & \cellcolor{gray!25}{$70.23^{+14.29}_{-13.03}$} & - \\
\hline
\multicolumn{2}{|c|}{Full Population} & 153 & $17.85^{+5.45}_{-4.30}$ & - & $0.82^{+0.90}_{-0.57}$ & - & $6.78^{+3.30}_{-3.79}$ & - & \cellcolor{gray!25}{$152.55^{+21.35}_{-18.75}$} & - \\
\hline
\end{tabular}

\caption{Values of case-specific local merger rate $R_0(\{\Theta_\text{GW}^i\})$, minimum time delay $t_d^\text{min}(\{\Theta_\text{GW}^i\})$ and power law index of DTD $d(\{\Theta_\text{GW}^i\})$ inferred in this analysis for high- and low-metallicity SFR. The values shown in \emph{grey} represent the estimated number of observable events $N_\text{est}$ for each case using the inferred values of $R_0(\{\Theta_\text{GW}^i\})$, $t_d^\text{min}(\{\Theta_\text{GW}^i\})$ and $d(\{\Theta_\text{GW}^i\})$ .}
\label{tab:estimates}

\end{table*}
}

\subsection{High-$Z$ SFR: $Z> 0.1\, Z_\odot$}
For the cases considered with high-metallicity-dependent SFR, we obtain results of the DTD for four classifications of GW source parameters denoted by $\Theta^i_{\rm GW}$ with $i \in \{1, 2, 3,4\}$. The corresponding corner plots 
on only the DTD parameters $t_d^{\rm min}$ and $d$ are shown in Fig. \ref{fig:corner-1}.  The results show that the minimum delay time is smaller for $m_1> M_{\rm PISN}$ ($t_d^\text{min}\sim0.85$ Gyr), irrespective of spin and mass ratio, than for the cases with $m_1< M_{\rm PISN}$ ($t_d^\text{min}\gtrsim3$ Gyr) (see Table \ref{tab:estimates} and (\textit{blue}, \textit{orange}) vs. (\textit{purple}, \textit{grey}) in Figs.\ \ref{fig:corner-1} \& \ref{fig:chieff_highZ}). Moreover, sources with high mass ratios tend to support longer minimum delay times in comparison to sources with lower mass ratios. Also, sources with $\chieff<0$ and masses lower than $M_{\rm PISN}$ exhibit the highest minimum delay-time ($t_d^\text{min}=4.41^{+1.05}_{-2.04}\,\text{Gyr}$). The power-law index ($d$) shows values away from one for all the cases $(d\sim7)$ except for $m_1< M_{\rm PISN}$ and $q<0.75$ ($d=3.03^{+6.33}_{-2.31}$), denoting that most of the cases show weak support for the vanilla case of flat-in-log distribution usually considered for scale-independent initial binary separation. We also show the merger rate results for only mass-ratio and effective spin in Fig. \ref{fig:q_chieff_highZ} in Appendix \ref{app:chi}, indicating the presence of sub-populations, without classifying separately in masses. 

These results, together with the corresponding high-metallicity SFR model, lead to an inference of the GW source merger rate for these cases. The figure indicates that the source-property-classified merger rate of BBHs exhibits different features and indicates the presence of three distinct classes of populations of BBH merger rate. Firstly, the sources with primary masses above and below the value of $M_{\rm PISN}$ show a different redshift evolution of the merger rate. Secondly, the local merger rate for the sources differs by about an order of magnitude depending on the source properties, denoting the absence of universality in the local merger rate of BBHs. Thirdly, the variation of the merger rate is stronger with variation in the mass-ratio than for effective spin, denoting that high-mass equal mass ratio systems are rare, in contrast to high-mass unequal mass ratio systems, denoting support for dynamical channels as a more plausible explanation to give rise to events with higher component masses with an unequal mass ratio. Our analysis indicates that the low redshift compact object events are primarily driven by the events with primary masses lower than M$_{\rm PISN}$, and events with primary masses above $M_{\rm PISN}$ dominate the high redshift part. This scenario flips at high redshifts, irrespective of their spin and mass ratio. We also show the inferred full-population merger rate distribution of BBHs $R(z)$ by the black line in Fig. \ref{fig:chieff_highZ}.

\subsection{Low-$Z$ SFR : $Z< 0.1\, Z_\odot$}
\begin{figure}
    \centering
    \includegraphics[scale=0.475]{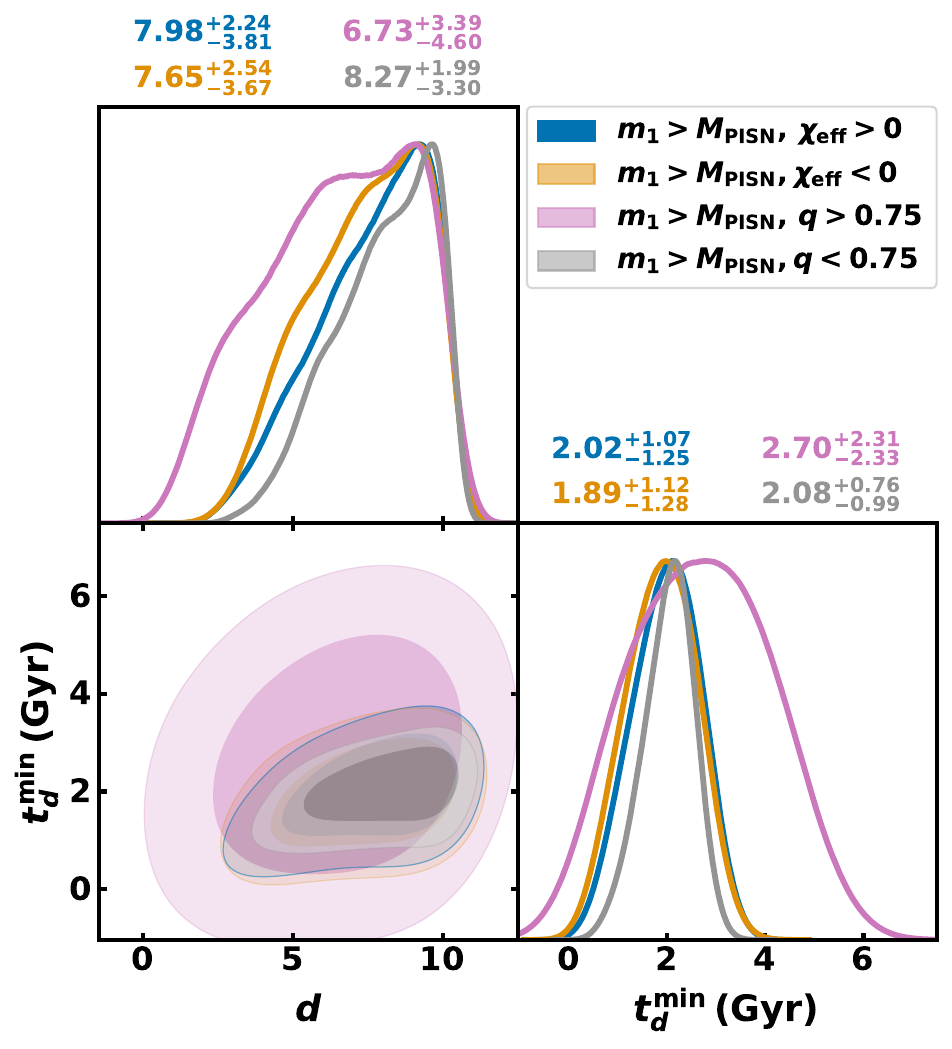}
    \caption{Inferred values of $t_d^\text{min}$ and $d$ with low-$Z$ SFR for the cases: $\chieff>0$ and $\chieff<0$ for $m_1>\text{M}_{\text PISN}$ ($\{\Theta^1_{\text{GW}}\}$), and $q>0.75$ and $q<0.75$ for $m_1>\text{M}_{\text PISN}$($\{\Theta^2_{\text{GW}}\}$).}\label{fig:corner-3}
\end{figure}

\begin{figure}
    \centering
    \includegraphics[scale=0.55]{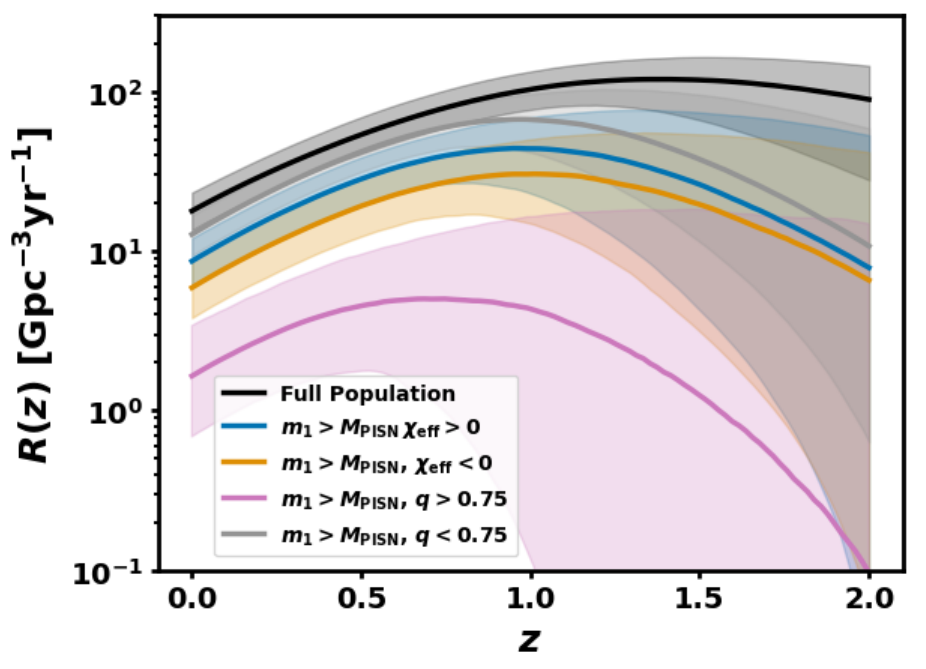}
    \caption{Redshift distributions inferred with low-$Z$ SFR for the cases: $\chieff>0$ and $\chieff<0$ for $m_1>\text{M}_{\text PISN}$ ($\{\Theta^1_{\text{GW}}\}$), and $q>0.75$ and $q<0.75$ for $m_1>\text{M}_{\text PISN}$($\{\Theta^2_{\text{GW}}\}$), along with the full-population distribution (shown in black).}\label{fig:lowZRGWz}
\end{figure}

For the analysis considered with low-metallicity-dependent SFR, we obtain results of the DTD for two classifications of GW source parameters denoted by $\Theta^i_{\rm GW}$ with $i \in \{1, 2\}$. The corresponding corner plot  
on only the DTD parameters $t_d^{\rm min}$ and $d$ is shown in Fig. \ref{fig:corner-3} and the corresponding BBH merger rate plot is shown in Fig. \ref{fig:lowZRGWz}. The corner plot indicates that the minimum delay time values are larger by $\sim1$Gyr in comparison to the cases with a high-metallicity SFR. This is expected as the low-metallicity SFR peaks at a higher redshift ($z\sim4$; see Fig.\ \ref{fig:sfrcomparison}), and in order to match the observed BBH source distribution, the minimum delay needs to be larger. Again, the power-law index ($d$) shows values away from one $(d\gtrsim7)$, ruling out possibilities of very large $t_d$. 

The overall behavior of the results on the BBH merger rate in comparison to the high-metallicity case remains the same. There is a clear indication of the existence of sub-populations in the DTD of the BBHs, indicating that the local merger rate depends on the BBH source properties and unequal mass-ratio sources or $\chieff>0$ are more likely than sources with equal mass-ratio or $\chieff<0$. 

\section{Discussion}\label{sec:discussion}
In the previous section, we showed that the current data exhibits sub-population in the DTD inference of the DTD based on their source properties. The summary of the inferred delay-time and local merger rate parameters is shown in Table \ref{tab:estimates}, along with the estimated number of events $N_{\rm est}$ from these inferred parameter values. The values of $N_{\rm est}$ for all the cases match well with the number of observed events in the catalog (third column in Table \ref{tab:estimates}), verifying that our inferred model agrees well with the observed population. Additional results are shown in Appendix \ref{app:chi}, which agrees with the conclusion that the BBH population of GWTC-4 exhibits source-dependent sub-populations of DTD. This new data-driven discovery can be connected to underlying formation channel scenarios based on different model-based predictions.  

Broadly, the three sub-group classifications identified in the previous analysis can be connected with some of the predictions from formation channels from isolated and dynamical formation \citep{Mandel:2015qlu, DiCarlo:2020lfa, Mapelli:2021taw, Boesky:2024wks}. The most dominant channel arises from the long delay time case for sources with primary masses less than M$_{\rm PISN}$, irrespective of their $\chieff$ and  $q$ values. This resembles the isolated BBH population. However, there exists a significant sub-population of merging BBHs with a smaller delay time and with masses above  M$_{\rm PISN}$ and unequal masses. This sub-population resembles the hierarchical formation channel of compact objects. An interesting sub-dominant population of sources exists, which is contributing towards a higher mass ratio and with component masses above M$_{\rm PISN}$ can originate from dynamical formation channels, as they predict a broad distribution in mass-ratio. Similarly, a sub-population of BBHs exists with $\chieff<0$ resembling a misaligned spin with orbital angular momentum, having a shorter time delay. This interesting population can originate from hierarchical mergers. The segregation of the BBH population based on the source-dependent DTD is able to connect to some of the key features of several population channels and is able to isolate the relative abundance of different populations present in the current LVK data.  

\section{conclusion}\label{sec:conclusion}
The detected population of BBH using the GW signal from the coalescing binaries has started to hint at their properties in the Universe. The existence of interesting correlations in the source parameters of these binaries and their evolution with cosmic time. In this study, we identify for the first time the existence of a sub-population of the DTD of the BBHs based on the black hole properties such as its mass, mass-ratio, and spin. The underlying population shows that BBHs that are heavier tend to merge more at a higher redshift than the sources with lighter ones. Moreover, there exists a critical sub-population of BBHs of higher masses with an unequal companion mass, which is driving the high redshift merger rate population present in the data, which closely resembles the hierarchical channel of BBH formation. This result also shows for the first-time the local merger rate of the BBH is very different for different sub-populations based on their sources properties, and there exists about an order of magnitude variation in the local merger rate of these sub-populations, with the highest arising from sources with their primary mass below M$_{\rm PISN}$ and the most sub-dominant arising from high primary mass systems with companion secondary mass comparable to the primary mass. 

The detected sub-population of DTD of the BBHs provides an interesting insight into the formation channels of these binaries and their relative abundances. One of the key findings of this work is the signature of different merger rates for different BBH source properties, classified based on their mass and spin. This indicates the breakdown of a universal merger rate of all BBHs, considering until now. Moreover, our results indicate the necessity to model the population of BBHs based on their sub-population in order to correctly capture their relative local merger rate. A recent analysis \citep{Cheng:2026bpc} has shown the existence of three sub-populations of merger rates for different mass values around 10 M$\odot$ and 35 M$_\odot$. In the future, with the help of more GW sources detected from the upcoming observations, this new approach will be able to strengthen the sub-population estimates and will also be able to shed light on the existence of any other rare sub-population in the data. The work establishes the presence of mass-dependent and spin-dependent DTD in the current LVK data, and the absence of any dependence is ruled out.   The future observations will reveal more about the relative abundance of these different sub-populations and their connections with different formation channels.

\begin{acknowledgments}
The authors are grateful to Upasana Das for carefully reviewing the manuscript and providing useful feedback to improve the draft. This work is a part of the ⟨Data$|$Theory⟩ Universe Lab, which is supported by the Department of Atomic Energy, Government of India. We acknowledge the support of the Department of Atomic Energy, Government of India, under Project Identification No. RTI 4012. This research is supported by the Prime Minister Early Career Research Award, Anusandhan National Research Foundation, Government of India. The authors are thankful to Martyna Chruslinska for providing the data files for metallicity-dependent star formation rate. We are also thankful for the computing resources provided by the ⟨Data$|$Theory⟩ Universe Lab. LIGO, funded by the U.S. National Science Foundation (NSF), and Virgo, supported by the French CNRS, Italian INFN, and Dutch Nikhef, along with contributions from Polish and Hungarian institutes. This collaborative effort is backed by the NSF’s LIGO Laboratory, a major facility fully funded by the National Science Foundation. 
\end{acknowledgments}

\begin{software}
    \ This work made use of the following software packages: \texttt{astropy} \citep{astropy:2013,astropy:2018,astropy:2022}, \texttt{Jupyter} \citep{2007CSE.....9c..21P,kluyver2016jupyter}, \texttt{matplotlib} \citep{Hunter:2007}, \texttt{numpy} \citep{numpy}, \texttt{pandas} \citep{mckinney-proc-scipy-2010,pandas-17229934}, \texttt{python} \citep{python}, \texttt{scipy} \citep{2020SciPy-NMeth,scipy-17467817}, \texttt{Bilby} \citep{bilby-paper,bilby-paper-2,Bilby-17533961}, \texttt{GetDist} \citep{Lewis:2019xzd,GetDist_20083735}, \texttt{Dynesty} \citep{Speagle:2019ivv}, \texttt{gwpopulation} \citep{Talbot2025}, \texttt{gwpopulation\_pipe} \citep{Talbot2021}.

    Software citation information aggregated using \texttt{\href{https://www.tomwagg.com/software-citation-station/}{The Software Citation Station}} \citep{software-citation-station-paper,software-citation-station-zenodo}.
\end{software}

\appendix

\section{Additional Results for different \protect\mbox{\ensuremath{\MakeLowercase{q}}} and $\chieff$ segregation}\label{app:chi}

\begin{figure}
    \centering
    \begin{minipage}{0.49\textwidth}
        \centering
        \includegraphics[scale=0.55]{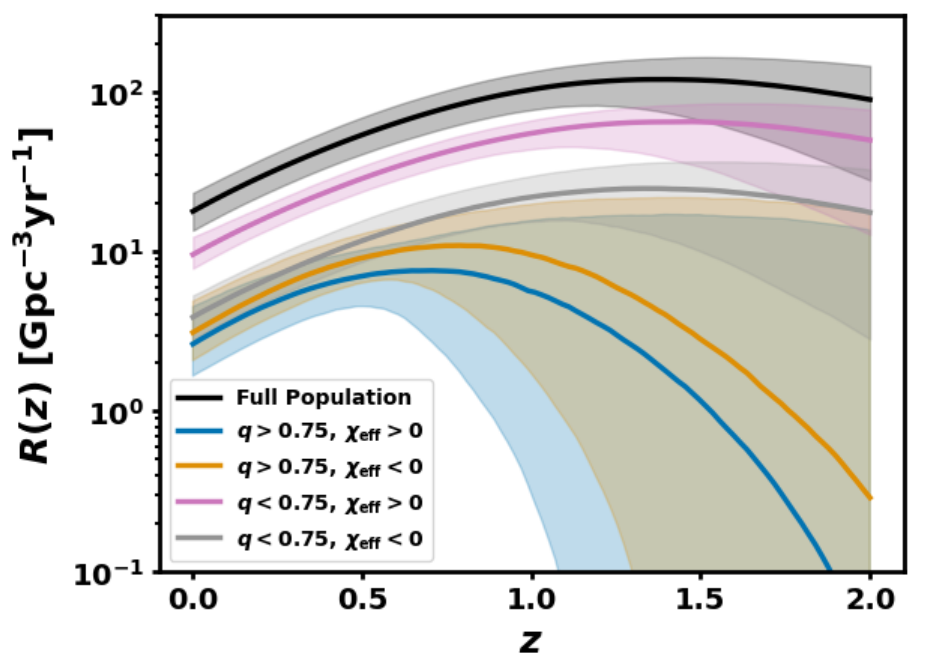}
    \end{minipage}
    \begin{minipage}{0.49\textwidth}
        \centering
        \includegraphics[scale=0.45]{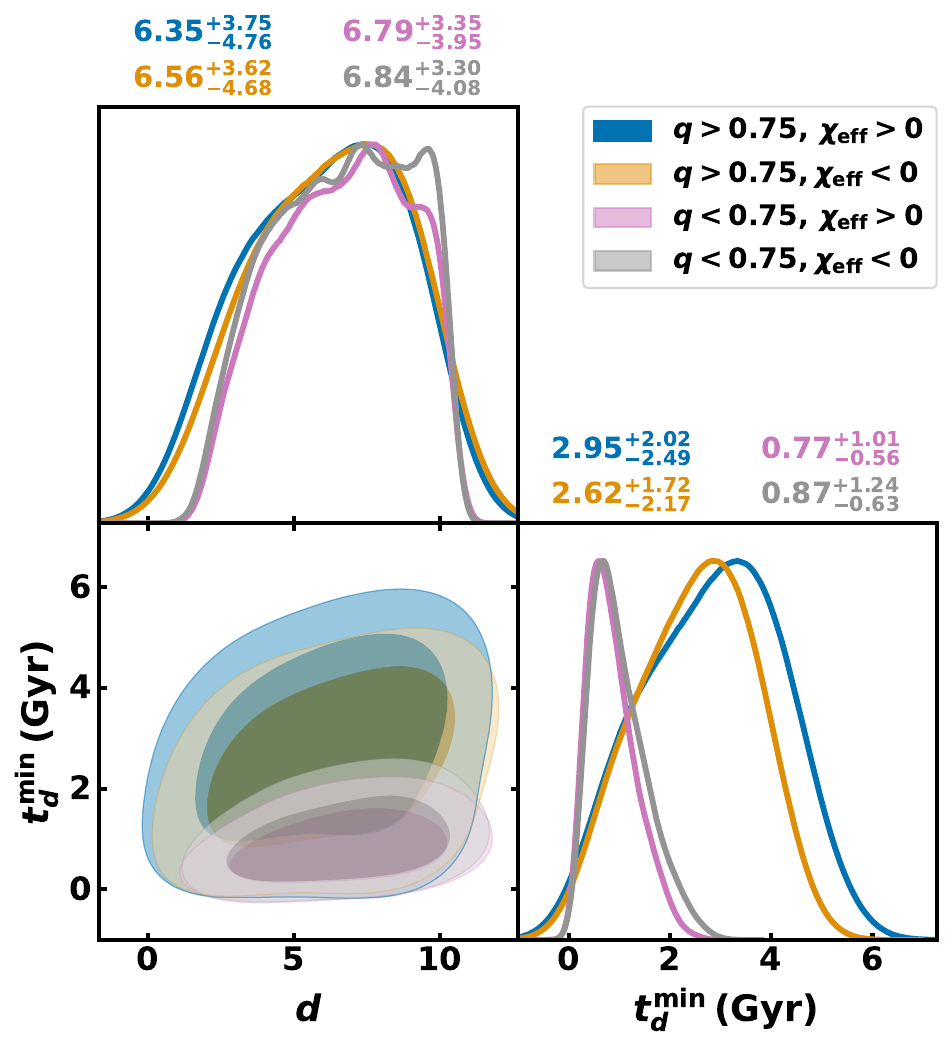}
    \end{minipage}
    \caption{Redshift distributions (\emph{Left}) and values of $t_d^\text{min}$ and $d$ (\emph{Right}) inferred with high-$Z$ SFR for the cases $(q>0.75, \chieff>0)$ (\emph{blue}), $(q>0.75, \chieff<0)$ (\emph{orange}), $(q<0.75, \chieff>0)$ (\emph{purple}) and $(q<0.75, \chieff<0)$ (\emph{grey}), with the full-population distribution in \emph{black}.}\label{fig:q_chieff_highZ}
\end{figure}

We also explore the variation of the DTD in $(q, \chieff)$ space and across the $\chieff$ range. We show the inferred redshift distributions and corresponding $t_d^\text{min}$ and $d$ values for the cases $(q>0.75, \chieff>0)$, $(q>0.75, \chieff<0)$, $(q<0.75, \chieff>0)$ and $(q<0.75, \chieff<0)$ using high-$Z$ SFR in Figs.\ \ref{fig:q_chieff_highZ}. It is evident from Fig.\ \ref{fig:q_chieff_highZ} that sources with near-equal mass ratios (\emph{blue, orange}) appear at lower redshift, while high redshifts are dominated by lower mass ratios (\emph{purple, grey}) with no strong dependence on $\chieff$. All results are quoted at 90\% credible intervals.

\begin{figure*}
    \centering
    \begin{minipage}{0.49\textwidth}
        \centering
        \includegraphics[scale=0.55]{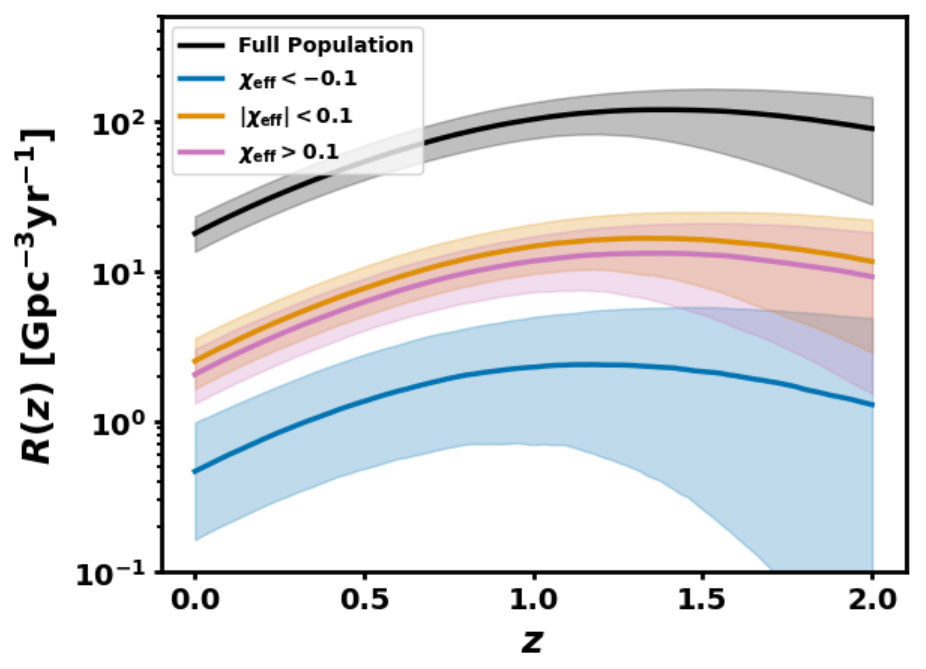}
    \end{minipage}
    \begin{minipage}{0.49\textwidth}
        \centering
        \includegraphics[scale=0.55]{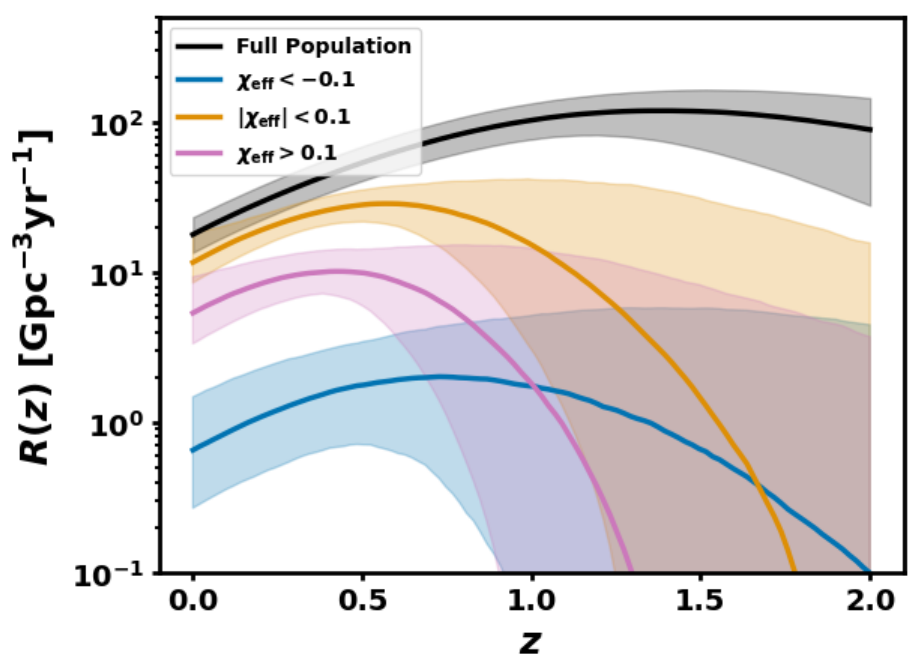}
    \end{minipage}
    \caption{Redshift distributions inferred for $\chieff < -0.1$ (\emph{blue}), $|\chieff|<0.1$ (\emph{orange}) and $\chieff > 0$ (\emph{purple}) for $m_1>\text{M}_\text{PISN}$ (\emph{Left}),  $m_1<\text{M}_\text{PISN}$ (\emph{Right}), and the full population case is shown in \emph{black}.}\label{fig:addplotRz}
\end{figure*}

\begin{figure*}
    \centering
    \begin{minipage}{0.49\textwidth}
        \centering
        \includegraphics[scale=0.475]{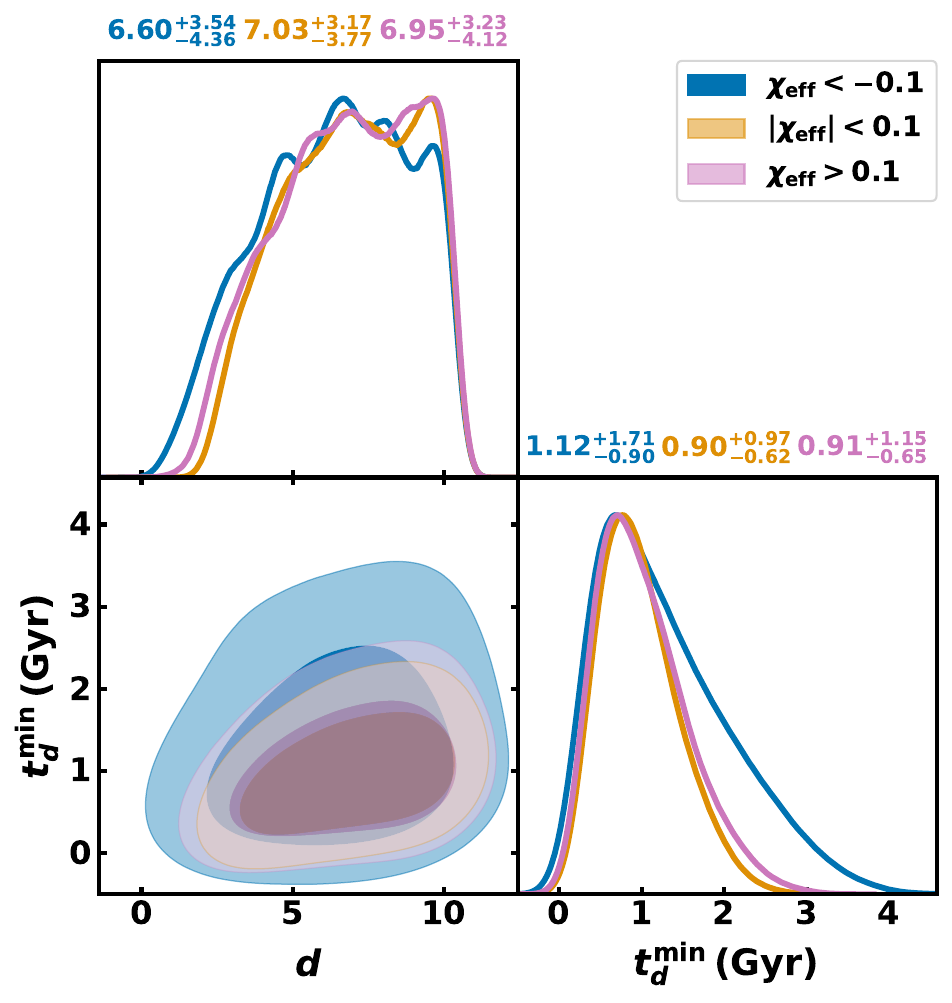}
    \end{minipage}
    \begin{minipage}{0.49\textwidth}
        \centering
        \includegraphics[scale=0.475]{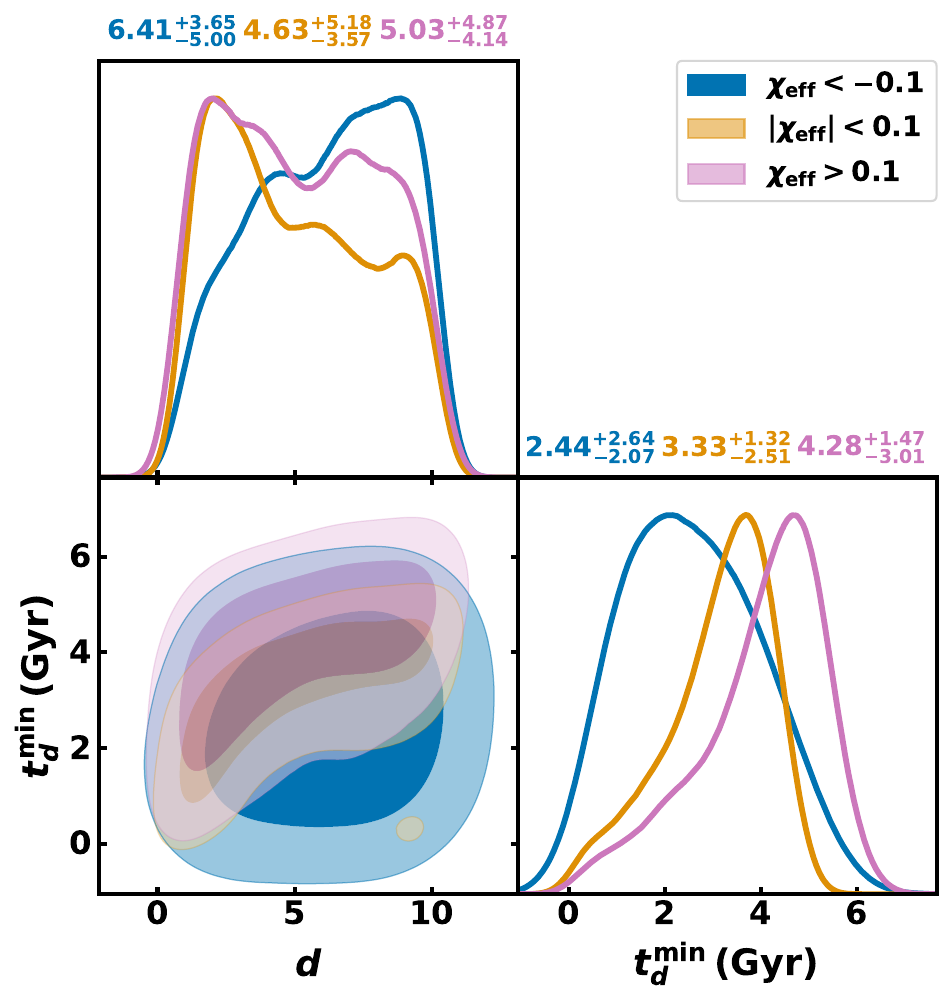}
    \end{minipage}
    \caption{Inferred values of $t_d^\text{min}$ and $d$ for $\chieff < -0.1$, $|\chieff|<0.1$ and $\chieff > 0$ for $m_1>\text{M}_\text{PISN}$ (\emph{Left}) and $m_1<\text{M}_\text{PISN}$ (\emph{Right}).}\label{fig:addplotcorner}
\end{figure*}

We also show the inferred redshift distributions and corresponding $t_d^\text{min}$ and $d$ values for the cases $\chieff<-0.1$, $|\chieff|<0.1$ and $\chieff>0.1$ for $m_1>\text{M}_\text{PISN}$ and $m_1>\text{M}_\text{PISN}$ in Figs.\ \ref{fig:addplotRz} \& \ref{fig:addplotcorner}. While the high-mass population shows negligible variation across the $\chieff$ bins, we find that $t_d^\text{min}$ increases monotonically with $\chieff$ for sources with $m_1<\text{M}_\text{PISN}$. Specifically, $\chieff<-0.1$ sources prefer the lowest minimum time delay $(t_d^\text{min}=2.44^{+2.64}_{-2.07}\,\text{Gyr})$, whereas sources with $\chieff>0.1$ prefer the highest minimum time delay $(t_d^\text{min}=4.28^{+1.47}_{-3.01}\,\text{Gyr})$ among these cases. Overall, these results are consistent with our findings in the main text and support our claim of the existence of subpopulations in the DTD of BBHs.
 
\bibliography{references}{}
\bibliographystyle{aasjournalv7}

\end{document}